\documentclass[journal]{IEEEtran}
\usepackage{amsmath,amsfonts}
\usepackage{algorithmic}
\usepackage{algorithm}
\usepackage{array}
\usepackage[caption=false,font=normalsize,labelfont=sf,textfont=sf]{subfig}
\usepackage{textcomp}
\usepackage{stfloats}
\usepackage{url}
\usepackage{verbatim}
\usepackage{graphicx}
\usepackage{cite}
\usepackage{siunitx}
\usepackage[version=4]{mhchem}
\usepackage{listings}
\usepackage{multicol}

\begin{document}

\title{Achieving Reliable and Repeatable Electrochemical Impedance Spectroscopy of Rechargeable Batteries at Extra-Low Frequencies}

\author{Christopher Dunn and Jonathan Scott,~\IEEEmembership{Life Senior Member,~IEEE}
\thanks{The authors are with the School of Engineering, University of Waikato, Private Bag 3105, Hamilton 3240, New Zealand.\\e-mail:~cjdunn@xtra.co.nz}
\thanks{Manuscript received February 9, 2022; revised ((date)), 2022.}}

\markboth{IEEE Transactions on Instrumentation and Measurement,~Vol.~XX, No.~X, Month~2022}%
{Shell \MakeLowercase{\textit{Dunn et al.}}: }

\IEEEpubid{0000--0000/00\$00.00~\copyright~2022 IEEE}

\maketitle

\begin{abstract}
There is a need for techniques for efficient and accurate measurement of the impedance of rechargeable batteries at extra-low frequencies (ELFs, typically below \SI{10}{\micro Hz}), as these reflect real usage and cycling patterns, and their importance in fractional battery circuit modeling is becoming increasingly apparent. Major impediments include the time required to perform such measurements, and `drift' in impedance values when measurements are taken from the same battery at different times. Moreover, commercial impedance analyzers are generally unable to measure at frequencies of the order of microhertz. We describe here our use of programmable two-quadrant power supplies to deliver multiple small signal measurement tones in the presence of large signal `working' currents, and our use of these data to generate impedance measurements with good precision and in reasonable time.
\end{abstract}


\section{Introduction}
\IEEEPARstart{E}{lectrochemical} impedance spectroscopy (EIS) is a powerful and non-destructive tool by which the properties and characteristics of a battery can be deduced by determining impedance over a wide frequency range~\cite{lasia_electrochemical_2014, choi_modeling_2020}. EIS is claimed to be a reliable indicator of state of charge (SoC), and can be used to predict state of health (SoH) \cite{ungurean_battery_2017}. Briefly, impedance $(Z)$ of an electrochemical system around some steady or quasi-steady state can be determined by \cite{zou_review_2018}:

\begin{enumerate}
\item Applying a sequence or set of small-signal alternating currents, expressed (for sinusoidal signals) as $ I(t) = |I|e^{j(\omega t + \phi_{I})} $;
\item Measuring the voltage response $ V(t) = |V|e^{j(\omega t + \phi_{V})} $;
\item Calculating $ Z(\omega) = \frac{|V|}{|I|}e^{j(\phi_{V} - \phi_{I})} $.
\end{enumerate}

$Z(\omega)$ is made up of real and imaginary parts:
\begin{center}
$ Z(\omega) = Z_{0}\cos(\phi) +  jZ_{0}\sin(\phi)$
\end{center}

where $ Z'_{real} = Z_{0}\cos(\phi) $, the resistance of the system, and $ Z''_{imag} = Z_{0}\sin(\phi) $, capacitance and/or inductance, representing energy storage~\cite{choi_modeling_2020}.

Impedance spectra can be represented as Bode plots, showing magnitude and phase shift across the measured frequency range, or as Nyquist (or Argand) plots, showing the real and imaginary parts of $Z(\omega)$ using cartesian coordinates.
The battery research community has traditionally tended to focus heavily on equivalent-circuit models (ECMs) based on Nyquist plots. These circuit models are considered key to prediction of characteristics such as SoC and SoH~\cite{cacciato_real-time_2017, choi_modeling_2020, westerhoff_analysis_2016, zou_review_2018, ungurean_battery_2017}. Nyquist plots are informative, but frequency information is implied rather than shown explicitly. Bode plots of magnitude and phase on the other hand contain all the necessary detail~\cite{lasia_definition_2014, westerhoff_analysis_2016, choi_modeling_2020}. This matters because the battery research community's preoccupation with the use of EIS to generate Nyquist plots to `look inside' batteries has resulted in a plethora of complicated and inconclusive ECMs, and has caused most workers to overlook a fundamental principle behind what is being measured and why.

\IEEEpubidadjcol

Mauracher and Karden noted as long ago as 1997 that rechargeable batteries ought to be measured and characterized at frequencies that reflect their usage patterns~\cite{mauracher_dynamic_1997}. These authors suggested that measuring down to \SI{10}{\micro Hz}, representing a period of approximately 28 hours, yielded useful information relating to diffusion outside the electrodes in a battery. Measurements taken at these extra-low frequencies (ELFs) might provide insights into model formulation~\cite{hasan_fractional_2016} or provide additional information to aid model fitting~\cite{ma_fractional_2016}. The use of ELFs has gone largely unnoticed by the research community, however, which is surprising because rechargeable batteries are often found in appliances that are charged daily, which corresponds to a cycling frequency of approximately \SI{11.6}{\micro Hz}. 
Other investigators report results obtained at frequencies no lower than the order of millihertz~\cite{guha_online_2018, chaoui_lyapunov-based_2015, do_impedance_2009, saha_prognostics_2009, cui_state_2018, jiang_electrochemical_2017, kim_-line_2020, yang_state--health_2020}. 
Some authors do not specify frequency ranges clearly, and most present only Nyquist plots.
The majority of commercial impedance meters have a lower limit of \SI{1}{mHz}.

Despite their acknowledgement of the potential usefulness of ELF measurement, Mauracher and Karden~\cite{mauracher_dynamic_1997} adopted the position that only frequencies down to \SI{50}{\micro Hz} were necessary (and subsequently modeled down to \SI{68}{\micro Hz}, although they measured down to \SI{6.8}{\micro Hz}) because electric vehicle excursions typically do not take more than 2 hours. In addition, they provided incomplete descriptions of how they obtained their data, which were sparse and noisy at lower frequencies.

\section{Difficulties Associated With Extra-Low Frequency Measurement}

Measuring in the ELF range is associated with a number of difficulties. These include the time needed to make the measurement. The minimum possible period for an EIS measurement is the reciprocal of its frequency. To minimize the time required, a single stimulus signal that includes all frequencies of interest can be used~\cite{scott_new_2019}. The magnitudes and phases of voltages and currents at these frequencies can be recovered by using for example a Fourier transform, and the complex impedance at each frequency calculated as described earlier. Various complex waveforms can be used to find impedances at multiple frequencies, for example a pseudo-random binary sequence (PRBS)~\cite{peinado_generation_2013, locorotondo_development_2021}, although there is a tradeoff between increasing numbers of frequencies and the signal-to-noise ratio (SNR) of the measurement. Alternatively, a small discrete set of tones might be used, in which case the stimulus is referred to as a multitone or multisine.

Researchers have also been faced with difficulties in finding equipment suitable for operation at ELFs. Most commercial impedance analyzers do not operate below \SI{1}{mHz}. A Solartron 1260A, which is capable of measuring down to \SI{10}{\micro Hz}, has been used for past work at this university, but was found to require great care when working with `wet' systems such as batteries~\cite{scott_new_2019}. Possible reasons for poor performance included drifting DC offset of the input signal and distortion associated with non-linear behavior outside the mid-range of a battery's SoC~\cite{scott_new_2019}. Furthermore, modern electrochemical theory accepts that batteries exhibit `fractional capacitance'~\cite{lasia_dispersion_2014}, and an ECM should therefore contain constant phase elements (CPEs) in order to embody battery characteristics~\cite{scott_new_2019, hasan_application_2016}. Fractional components are associated with long decay tails after a stimulus is applied or removed~\cite{scott_compact_2013, hasan_extending_2020}, and these tails are thought to account for the problems encountered when measuring `wet' systems with the Solartron.

\begin{figure}[htbp]
\centering
\includegraphics[scale=0.62, trim = 100 250 100 250, clip]{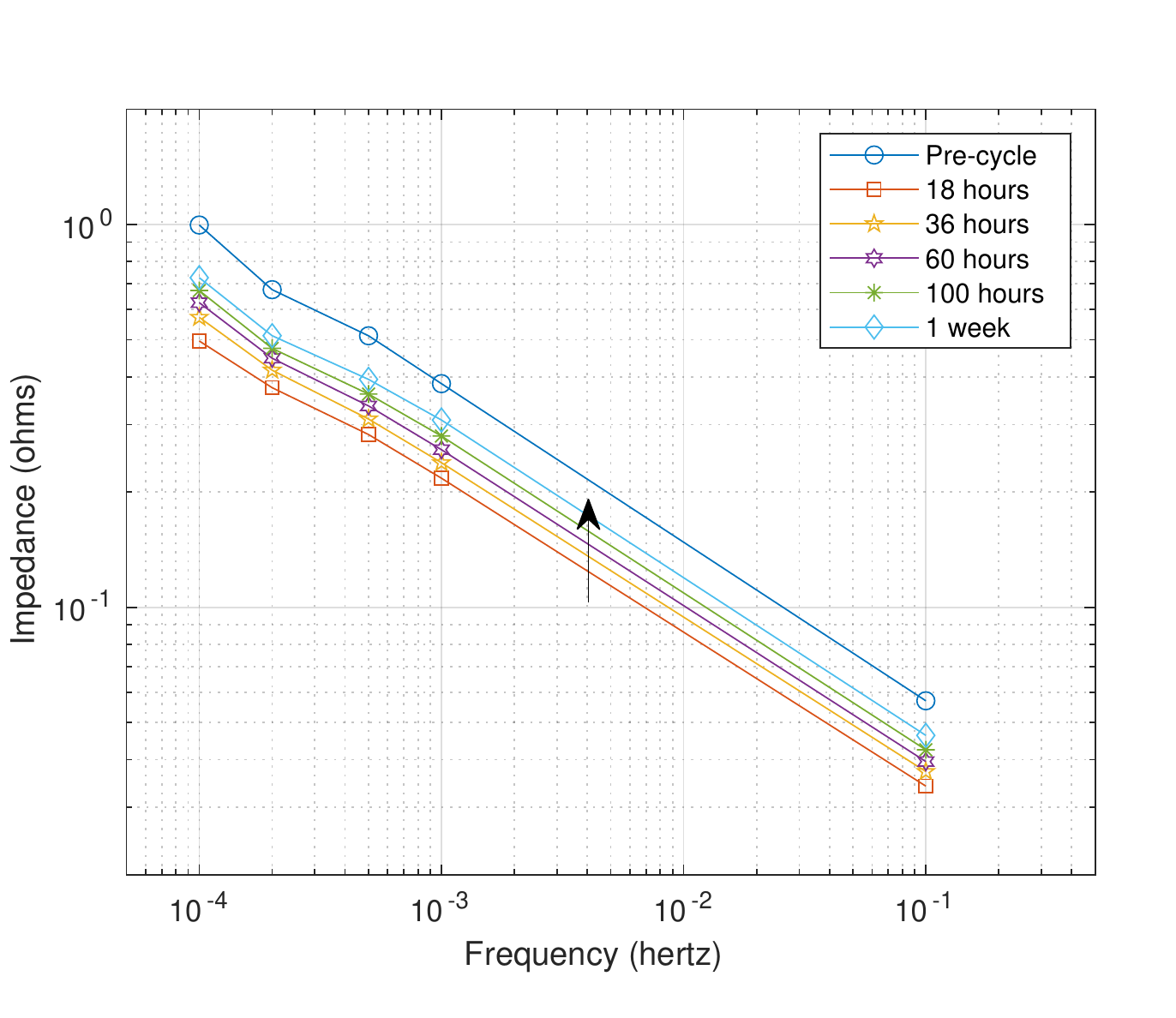}
\caption{Impedance (magnitude) of a \SI{30}{Ah} AGM lead-acid battery at five frequencies before and at various times after a charge-discharge cycle at \SI{5}{A}. The arrow shows the impedance drift over time as the battery equilibrates.}
\label{AGM_PbA.pdf}
\end{figure}

In addition, the apparent impedance of a battery appears to depend on factors such as SoC, charging or discharging, any superimposed DC current, short-term history and homogeneity of the electrolyte~\cite{budde-meiwes_influence_2011}. Figure~\ref{AGM_PbA.pdf} shows EIS data obtained at five frequencies from a near-new \SI{30}{Ah} absorbent glass mat (AGM) lead-acid battery of the type typically found in light traction and power backup applications. Measurements were made before and at several times after cycling the battery and returning it to 80\% SoC. Note that the small-signal impedance of the battery is halved by cycling, with over a week being required for the battery to re-equilibrate and return to its original state.
Other reported transient phenomena include the `coup de fouet' (`crack of the whip'), a sudden drop in voltage with subsequent recovery when a fully charged lead-acid battery is discharged~\cite{pascoe_behaviour_2002}.

\begin{figure}[htbp]
\centering
\includegraphics[scale=0.22, trim = 10 150 10 150, clip]{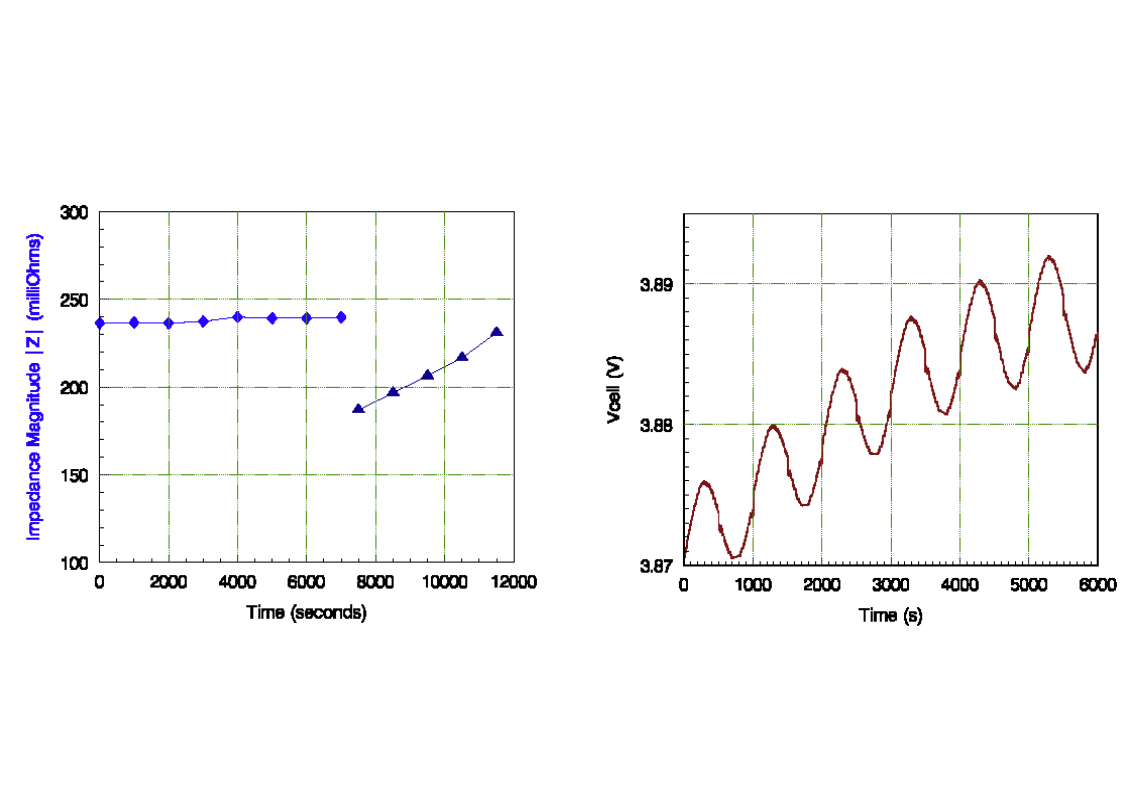}
\caption{Recovery of a lithium nickel cobalt (NCO) battery after cycling. Left: impedance magnitude at \SI{0.001}{Hz} before and after a charge-discharge cycle at \SI{2}{A}; Right: voltage waveform in the time domain for the impedance measurement immediately after the cycle.}
\label{fbq_fig2.png}
\end{figure}

Impedance drift is seen in other battery chemistries, although the re-equilibration process may be considerably faster and not as easy to observe. Figure~\ref{fbq_fig2.png} shows impedance at a single frequency (\SI{1}{mHz}) of a lithium-ion battery before and after cycling.

Thus, batteries must apparently be allowed to rest after cycling before their impedances are measured. However, a real-world battery will not be operating under laboratory conditions: the cycling currents mimic the types of large signals that batteries experience when they are in use, and any realistic model or battery management system (BMS) must take account of this. For accurate modeling and SoH prediction, measurements need to be stabilized using a method that could eventually find its way into a BMS.

\section{Current Magnitude and Charge Density}

\begin{figure}[htbp]
\centering
\includegraphics[scale=0.62, trim = 100 250 100 250, clip]{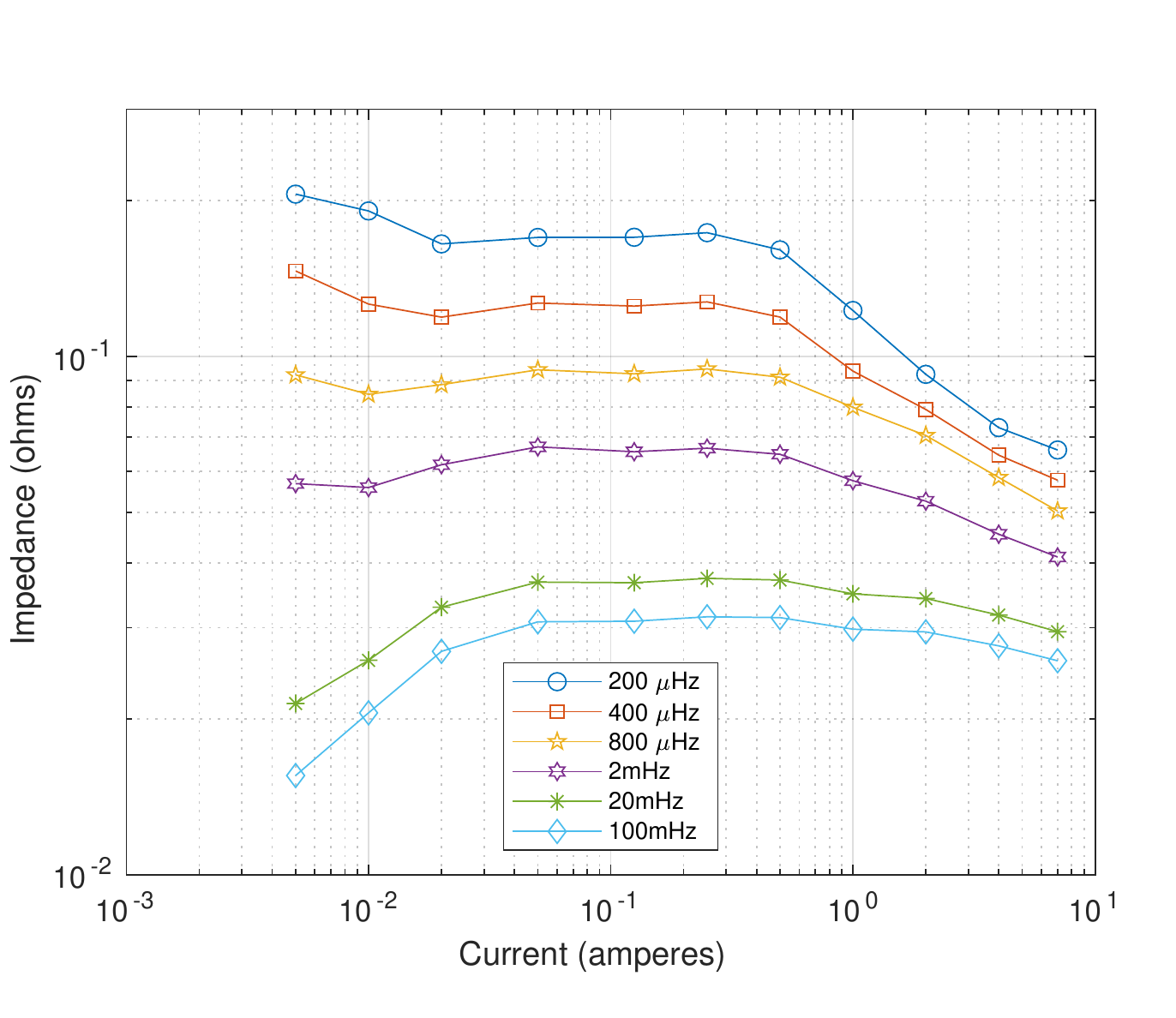}
\caption{Impedance of a \SI{4}{Ah} \ce{LiFePO4} battery plotted against current at six frequencies. Measurements taken on a Keithley 2460A four-quadrant precision source controlled by a Raspberry Pi running custom software \cite{farrow_characterisation_2020}.}
\label{ziPlotPub.pdf}
\end{figure}

\begin{figure}[htbp]
\centering
\includegraphics[scale=0.62, trim = 100 250 100 250, clip]{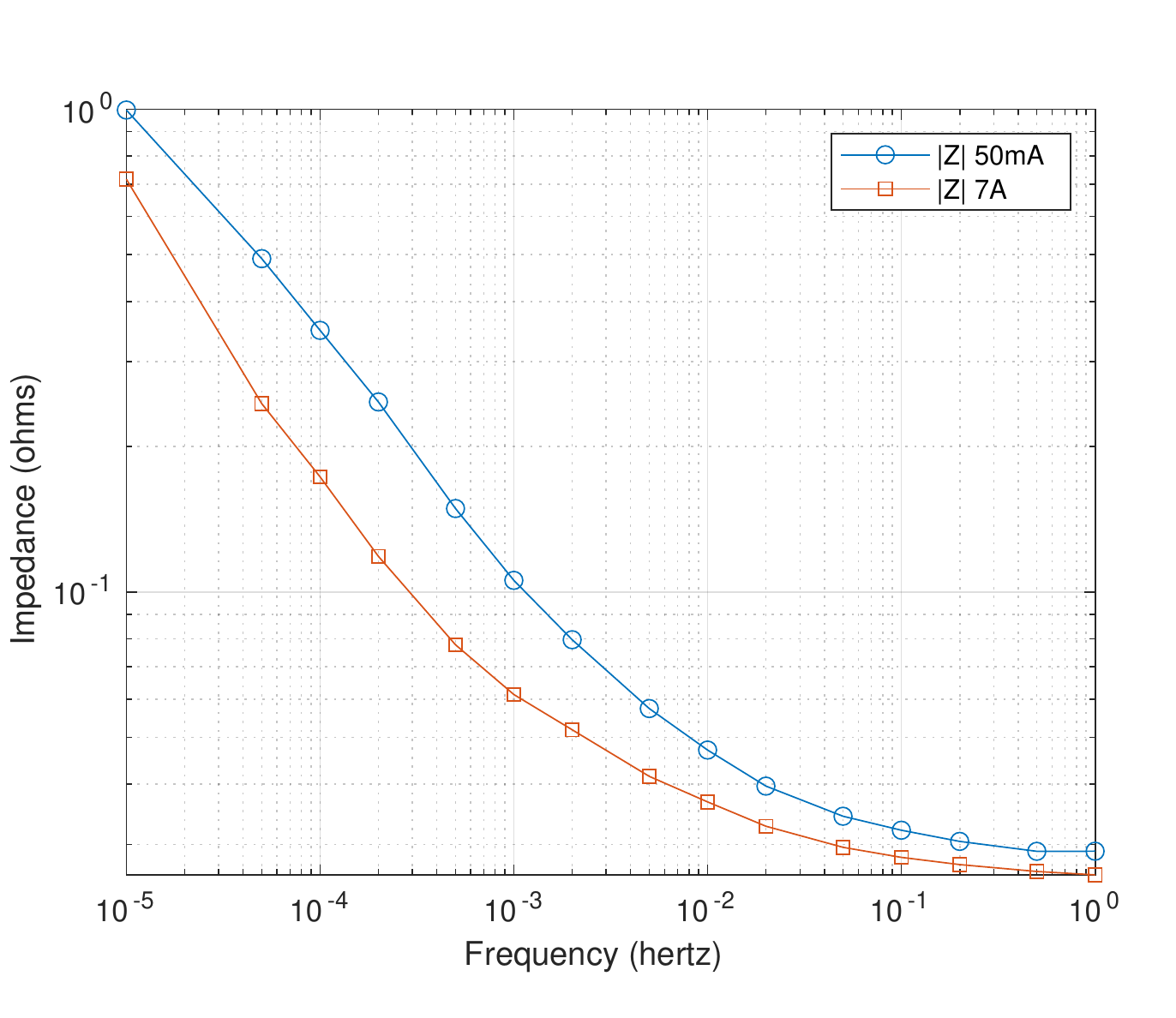}
\caption{Impedance versus frequency for a \SI{4}{Ah} \ce{LiFePO4} battery at low and high currents. Full frequency sweeps from \SI{10}{\micro Hz} to \SI{1}{Hz}; \SI{50}{mA} and \si{7}{A}.}
\label{LK2_Z_50mA_7A.pdf}
\end{figure}

Evidently, charge density or rates of current flow must be taken into account when measuring impedance and parameterizing models. Apparent impedance decreases when higher currents are passed, an effect that is exaggerated at low frequencies. Figure~\ref{ziPlotPub.pdf} shows data from an exploratory experiment in which $|Z|$ was measured using a multisine run at maximum current ($I_{max}$) values ranging from \SI{20}{mA} to \SI{7}{A} on a lithium iron phosphate battery rated at \SI{4}{Ah}. Frequencies ranged from \SI{200}{\micro Hz} to \SI{100}{mHz}. Measurements from \SI{50}{mA} to \SI{200}{mA} (approximately C/50) behaved like `small' signals. At higher frequencies, increasing current had less effect, but although the \SI{7}{A} maximum current had a marked effect on the low frequency traces, it did not `pull' the impedance down to a high current asymptote, which suggests some permanent resistive mechanism that cannot be overcome. This is more apparent in Figure~\ref{LK2_Z_50mA_7A.pdf}, which shows full frequency sweeps carried out with $ I_{max} $ set at \SI{50}{mA} and \SI{7}{A} after cycling.

Unfortunately, it is not possible to draw substantial current from a battery for prolonged periods because the battery will run flat, a major difficulty encountered by Budde-Meiwes et al.~\cite{budde-meiwes_influence_2011}. As described by Scott and Hasan~\cite{scott_new_2019}, the peak amplitude of charge delivered is dependent on the frequency of the signal. As frequency decreases, the current stimulus must be dropped significantly to prevent the charge excursion flattening or overcharging the battery. Budde-Meiwes et al. abandoned ELF measurements because it is not possible to make a reliable measurement of battery impedance below \SI{1}{mHz} without running for periods well in excess of 1000~seconds. Overall, the twin aims of measuring at ELFs while the battery under test is delivering significant current cannot be met with currently available techniques. 
In this manuscript we describe two methods permitting ELF measurements while the battery is delivering or receiving substantial current or charge without going flat or overcharging.

\section{New Methods For Battery Impedance Measurement}

All data presented here were measured at \SI{22}{\degree C} with two-quadrant precision sources using four-wire connections controlled via GPIB interfaces by Raspberry Pi 4 computers running custom software written in C. The software utilities are bz3p66 and bzdcp66 (respectively, battery impedance with triphasic pulses or added DC component on our `66332' hardware; see Appendix). These command line programs minimize the risk of interference between the small signal measurement signal and the large `working' current signal by using time division or frequency division multiplexing, respectively. Both programs make multitone impedance measurements by sourcing and sinking current and recording times, currents and voltages in three-column `.tvi' data files (timestamp, voltage, current). Default frequency ranges are 1-2-5 sequences between minimum and maximum specified values, but any set of frequencies is possible. Complex impedance is calculated from the .tvi data by a discrete Fourier transform performed by `dftp', a program based on software originally described by Scott and Parker~\cite{scott_distortion_1995} for use with SPICE. bz3p66 and bzdcp66 are usually called by scripts setting the parameters for each run.

The battery used in the new ELF experiments is a brand new lithium titanate (LTO) \SI{40}{Ah} \SI{2.3}{V} 66160H cell.

\begin{figure}[htbp]
\includegraphics[scale=0.62, trim = 100 250 100 285, clip]{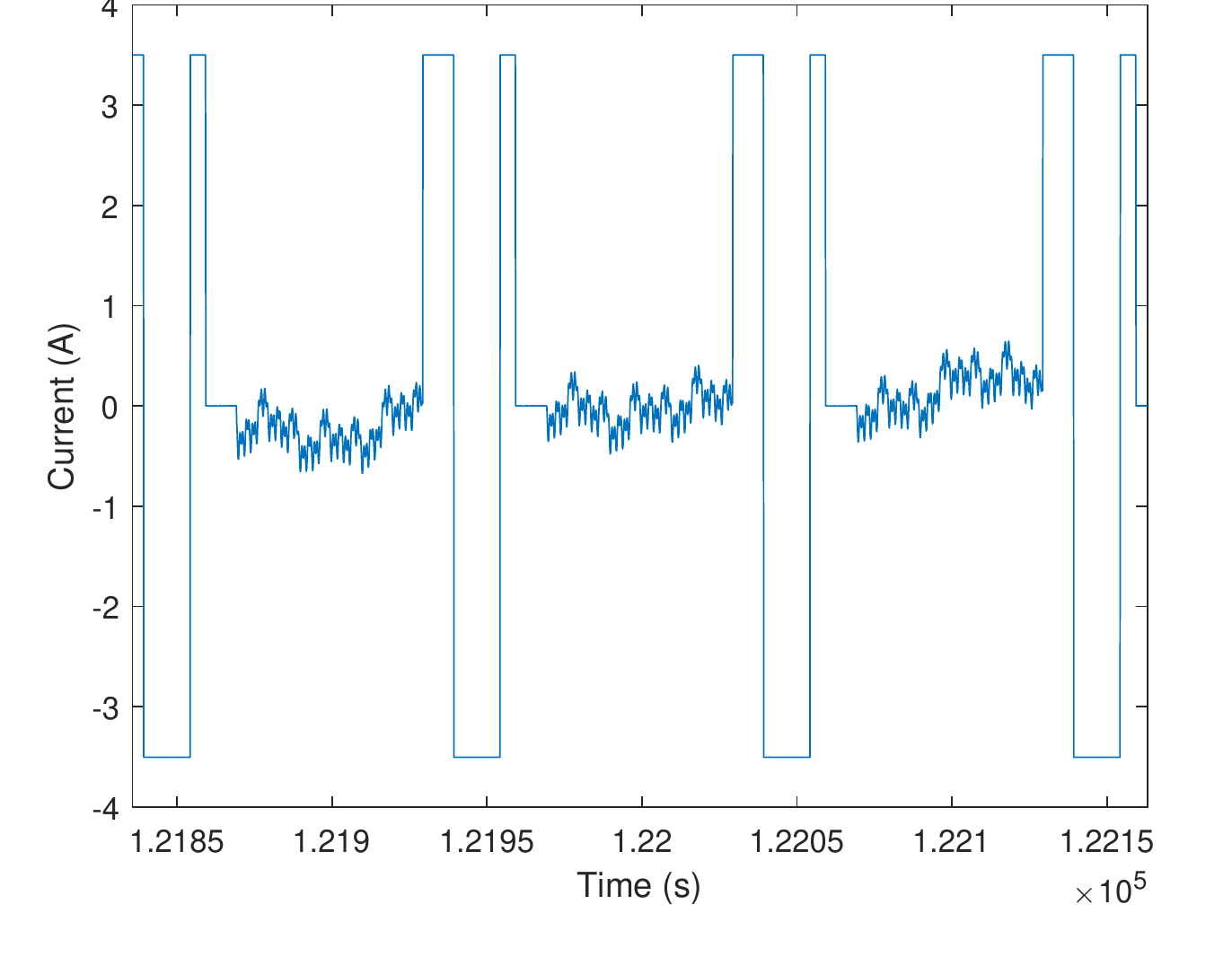}
\caption{Time domain plot: bz3p66 on LTO, 30-second \text{$\pm$}\SI{3.5}{A} triphasic pulse every minute. Note the 10-second `rest' after the end of each pulse to allow for settling of the long voltage tail that characterizes fractional devices, and the small signal measurement tones between the pulses.}
\label{bz3p_Freqs_30s_q1min.ptvi.pdf}
\end{figure}

\subsection{Time Division Multiplexing: bz3p66}

bz3p66 alternately subjects the battery to a small-signal measurement multitone and a `working' stimulus signal that reflects a normal usage current. Interference is minimized by reducing as far as possible the change in terminal voltage that is seen following the application of a large working signal (i.e. minimizing the artefact or `pulse tail' that characterizes fractional elements as described earlier~\cite{hasan_extending_2020}). The working stimulus is a triphasic pulse, a stimulus believed from empirical work with human-implanted electrodes to minimize artefact tails in implantable medical devices~\cite{scott_compact_2013}. Some mathematics confirms that the optimal pulse in a fractional-derivative system consists of three equal-amplitude, alternating-polarity pulses in a 2-3-1 duration sequence (Figure~\ref{bz3p_Freqs_30s_q1min.ptvi.pdf}).

bz3p66 outputs two time domain files, one that includes the triphasic pulse (`.ptvi') and one with the pulse removed. The latter is the .tvi file that is used for further analysis.

The LTO battery was cycled once using a constant current-constant voltage (CCCV) regime. Maximum ($ V_{max} $) and minimum ($ V_{min} $) CV voltages were set at \SI{2.7}{V} and \SI{1.8}{V}, respectively; the charge and discharge CC currents were \SI{5}{A}. CV phases were ended at $ \pm $\SI{1}{A}. The battery was rested after cycling for 10 minutes at a final SoC of 65\%. Following this, bz3p66 was run four times with working $ \pm $\SI{3.5}{A} current pulses as follows:

\begin{enumerate}
\item{30-second pulse every 4 minutes;}
\item{30-second pulse every 2 minutes;}
\item{30-second pulse every minute (see Figure~\ref{bz3p_Freqs_30s_q1min.ptvi.pdf});}
\item{60-second pulse every 2 minutes.}
\end{enumerate}

Measurement tones were 1-2-5 sequences from \SI{10}{\micro Hz} (three cycles) to \SI{1}{Hz}. The maximum total charge that could be sourced or sunk ($ dQ_{max} $) was set at \SI{5}{Ah} (0.125C). This charge is distributed across all measurement tones.

\begin{figure}[htbp]
\includegraphics[scale=0.62, trim = 100 250 100 285, clip]{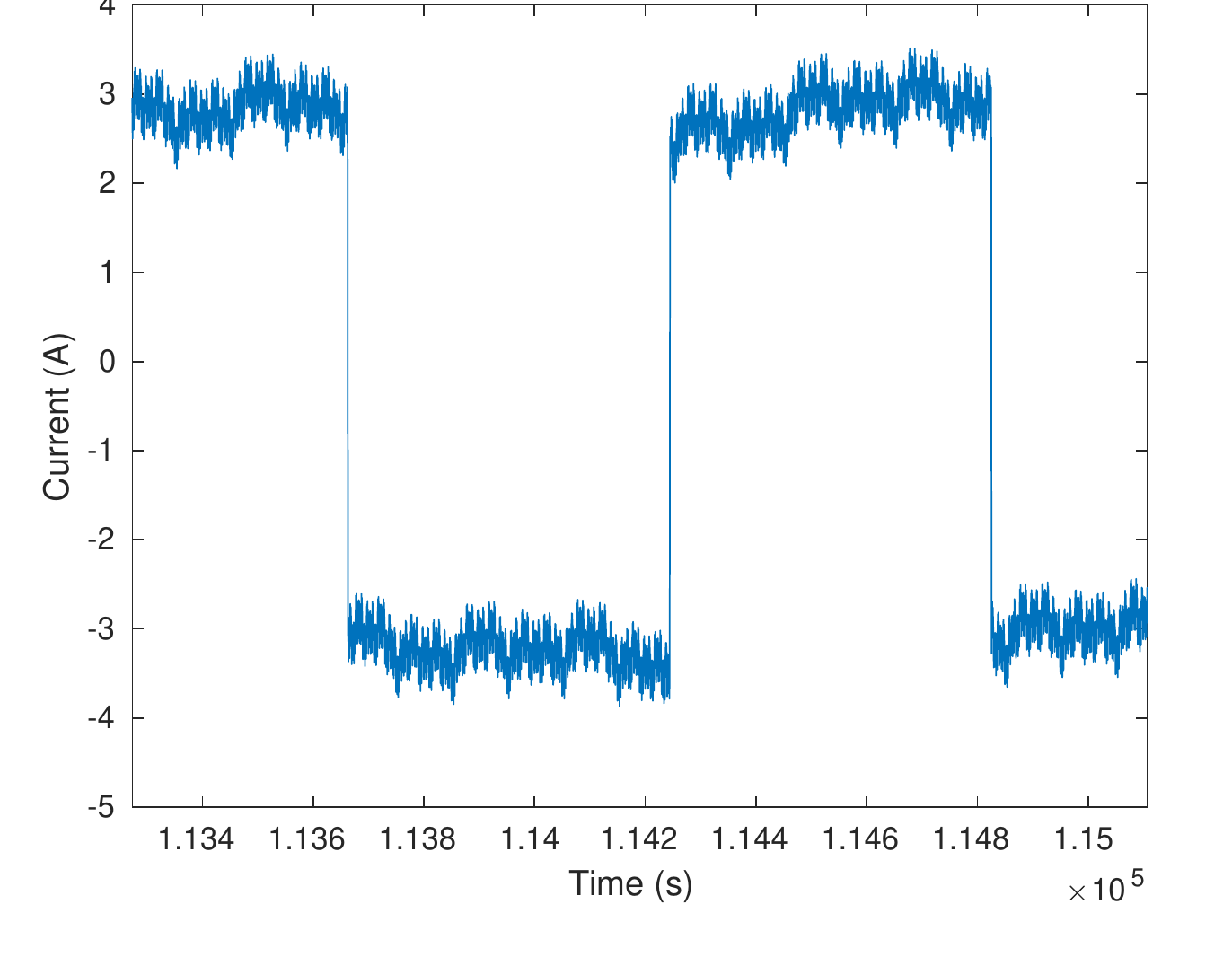}
\caption{Time domain plot: bzdcp66 on LTO, $\pm$\SI{3}{A} \SI{860}{\micro Hz} square wave. Note offset measurement tones.}
\label{tviPlot_bzdc_3A_1.pdf}
\end{figure}

\subsection{Frequency Division Multiplexing: bzdcp66}

bzdcp66 performs battery measurements by exposing the battery to a small signal multitone and a working stimulus signal simultaneously. This is accomplished by superimposing the EIS multitone on a periodic square wave which aims to mimic the effect of a working DC signal (Figure~\ref{tviPlot_bzdc_3A_1.pdf}).

The battery was cycled to 65\% SoC as for bz3p66 and then subjected to sets of consecutive bzdcp66 runs as follows:

\begin{enumerate}
\item{three runs with \SI{3}{A} square wave ($ I_{dc} $) at a frequency of \SI{860}{\micro Hz} ($ f_{dc} $), i.e. a period of approximately 20 minutes;}
\item{three runs with $ I_{dc} $ \SI{35}{mA} with the same $ f_{dc} $ (i.e. negligible working current signal);}
\item{cycle again;}
\item{repeat (1) and (2).}
\end{enumerate}

Measurement tones were 1-2-5 sequences from \SI{20}{\micro Hz} (three cycles) to \SI{2}{Hz}. $ dQ_{max} $ was \SI{5}{Ah}, distributed across all measurement tones as with bz3p66.

The working and EIS signals for bzdcp66 are selected so that their spectra do not overlap. This is done by choosing EIS frequencies that are multiples of $ f_0 $, the lowest frequency, and powers of 2 or 5:
\begin{equation}
f_{test} = 2^p \times 5^q \times f_0
\end{equation}
where $ f_{test} $ is a test frequency, and $ p $ and $ q $ are integers. Thus, each frequency in the multisine has the prime factors 2 and 5 only in its multiple of $ f_0 $. The square wave is then given a frequency that is a multiple of $ f_0 $ but that does not have 2 or 5 as prime factors, for example:
\begin{equation}
f_{dc} = 7^i \times 11^j \times 13^k \times 17^l \times f_0
\end{equation}
where $ i $, $ j $, $ k $ and $ l $ are integers. This suggests that $ f_{dc} $ might be a multiple of 7, 11, 13, 17, etc. 

As a final check, all odd multiples of $ f_{dc} $ can be compared with each multisine tone, and a larger factor chosen for $ f_{dc} $ if there are `near miss' differences that might pollute adjacent tones in a windowed Fourier transform. For these experiments, \SI{860}{\micro Hz} was used as 860 is a multiple of a prime number (43) and does not have 2 or 5 as prime factors.

\section{Results}

\subsection{Time Division Multiplexing: bz3p66}

\begin{figure}
\includegraphics[scale=0.62, trim =100 250 80 250, clip]{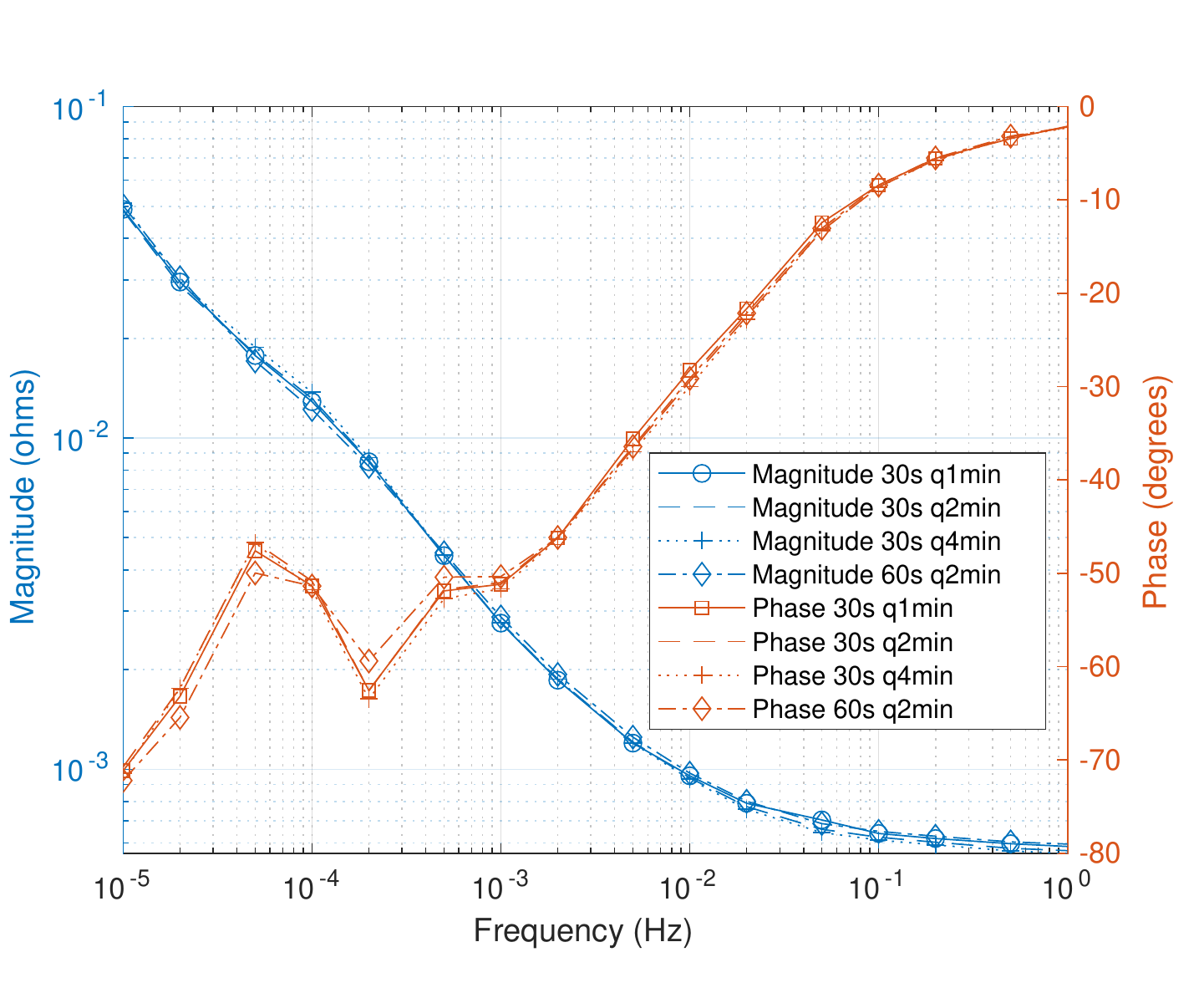}
\caption{bz3p66 on LTO battery: impedance magnitude and phase of four runs with differing triphasic pulse frequencies and durations. 30s q1min = 30-second triphasic pulse repeated every minute, 30s q2min = 30-second pulse repeated every 2 minutes, etc.}
\label{bz3p_fmp_All.pdf}
\end{figure}

Impedance magnitude and phase curves obtained from the Fourier-transformed .tvi data generated by bz3p66 are shown in Figure~\ref{bz3p_fmp_All.pdf}. Varying the duration and/or frequency of the triphasic pulse made little or no difference to magnitude and phase profiles, with consistent disturbance of the phase response around \SI{0.1}{mHz}. The reason for this is unclear.

While detailed interpretation of the impedance data is beyond the scope of this paper, different regions of the Bode plot reveal various aspects of the battery's characteristics and condition that can be represented in an ECM. Briefly, the high frequency section of the magnitude plot approaches a horizontal asymptote that represents the series resistance ($ R_s $) of the battery, while the region below \SI{1}{mHz} is dominated by a CPE. Other behaviors of the magnitude and phase plots reveal hints as to other circuit elements that might be appropriate. For example, the rounded `knee' between the $ R_s $ and first CPE regions suggests a second fractional series element. Interested readers are referred elsewhere for more information~\cite{scott_new_2019, hasan_extending_2020, poihipi_distinguishability_2021}.

\subsection{Frequency Division Multiplexing: bzdcp66}

\begin{figure}[htbp]
\includegraphics[scale=0.62, trim =100 250 80 250, clip]{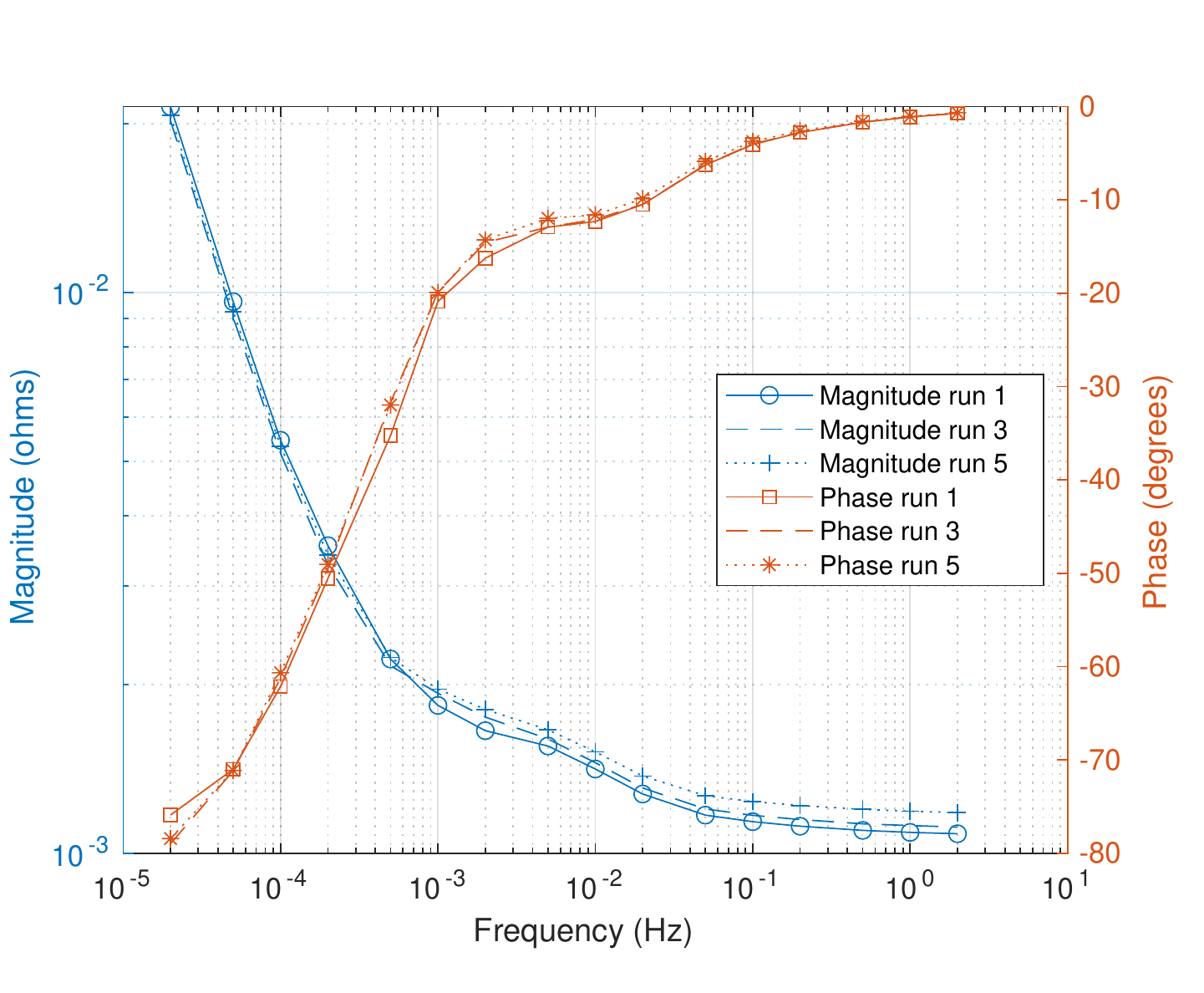}
\caption{Impedance magnitude and phase plots: bzdcp66 runs 1, 3 and 5 on LTO battery with \SI{3}{A} square wave.}
\label{fmpPlot_135_3A.pdf}
\end{figure}

\begin{figure}[htbp]
\includegraphics[scale=0.62, trim =100 250 80 250, clip]{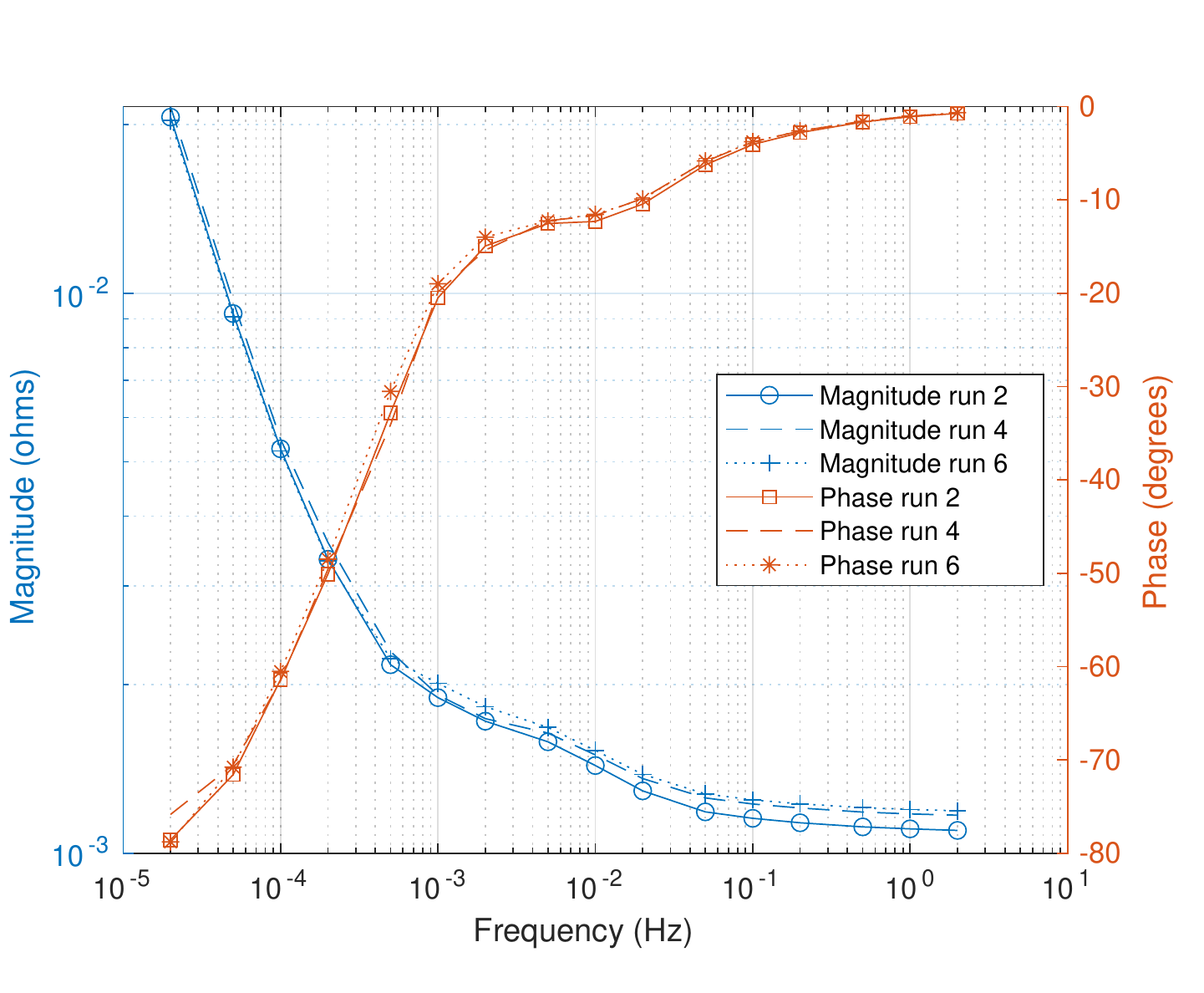}
\caption{Impedance magnitude and phase plots: bzdcp66 runs 2, 4 and 6 on LTO battery with \SI{3}{A} square wave.}
\label{fmpPlot_246_3A.pdf}
\end{figure}

Figures~\ref{fmpPlot_135_3A.pdf} and \ref{fmpPlot_246_3A.pdf} show impedance magnitude and phase curves overlaid from repeated runs (three plots per chart to maintain clarity) of bzdcp66 with a \SI{3}{A} working square wave. Magnitude and phase appeared consistent and repeatable between runs, whether interspersed with additional cycling or not, and the low frequency phase response did not show the disturbance exhibited repeatedly by bz3p66.

\begin{figure}[htbp]
\includegraphics[scale=0.62, trim =100 250 80 250, clip]{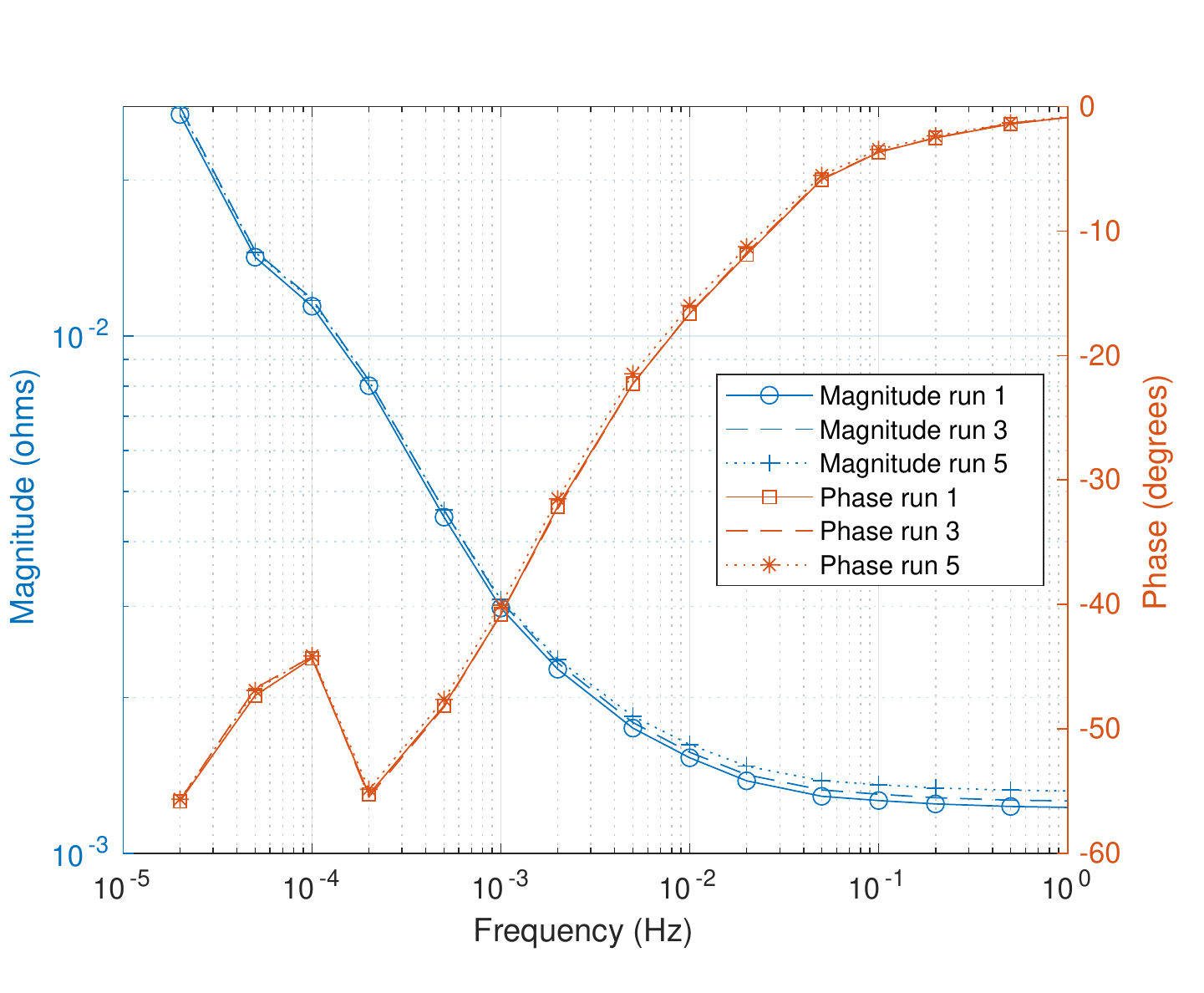}
\caption{Impedance magnitude and phase plots: bzdcp66 runs 1, 3 and 5 on LTO battery with \SI{35}{mA} square wave.}
\label{fmpPlot_135_35mA.pdf}
\end{figure}

\begin{figure}[htbp]
\includegraphics[scale=0.62, trim =100 250 80 250, clip]{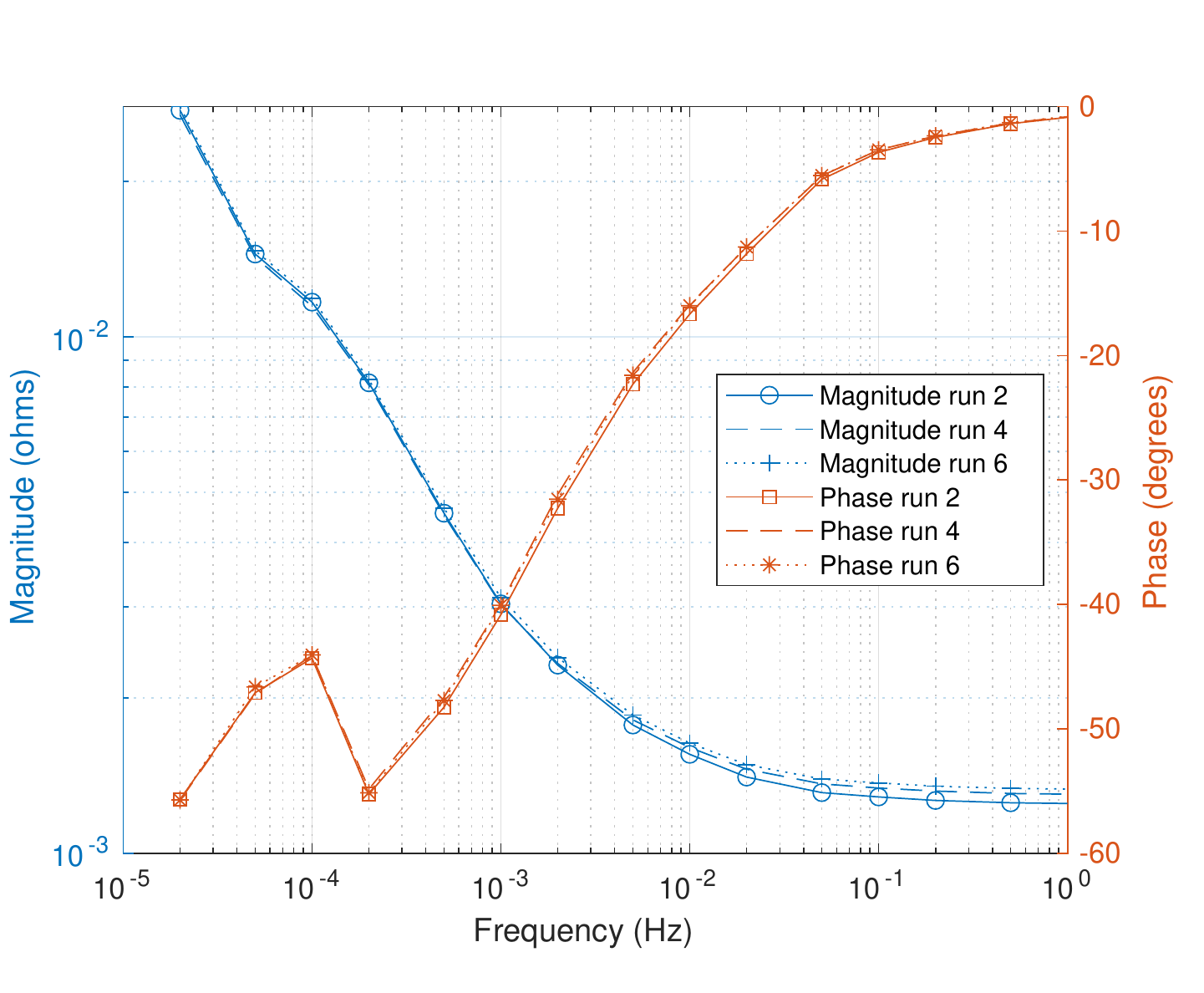}
\caption{Impedance magnitude and phase plots: bzdcp66 runs 2, 4 and 6 on LTO battery with \SI{35}{mA} square wave.}
\label{fmpPlot_246_35mA.pdf}
\end{figure}

Similar consistency between runs was noted when bzdcp66 was run with a negligible (\SI{35}{mA}) square wave to mimic a small signal measurement multitone with no working current (Figures~\ref{fmpPlot_135_35mA.pdf} and \ref{fmpPlot_246_35mA.pdf}). No impedance drift due to re-equilibration after cycling as described earlier (Figures~\ref{AGM_PbA.pdf} and \ref{fbq_fig2.png}) was evident. These results appear to confirm earlier observations of very rapid re-equilibration of batteries based on lithium chemistries (Figure~\ref{fbq_fig2.png}). Interestingly, the phase irregularity around \SI{0.1}{mHz} seen with time division multiplexing reappeared in the presence of negligible $ I_{dc} $.

\subsection{Comparison of Time Division and Frequency Division Multiplexing: bz3p66 vs bzdcp66}

\begin{figure}[htbp]
\includegraphics[scale=0.62, trim =100 250 80 250, clip]{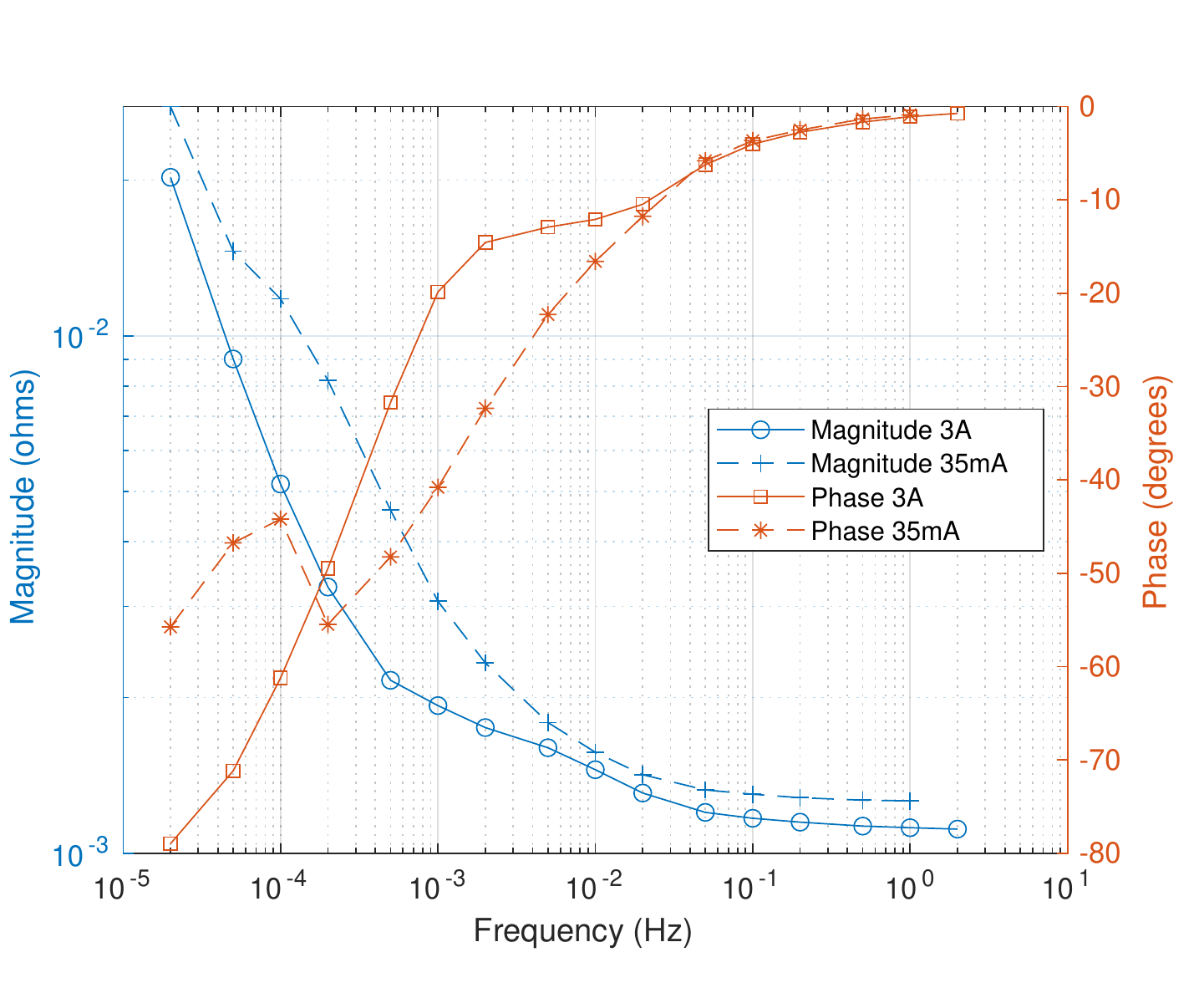}
\caption{Impedance magnitude and phase plots: representative bzdcp66 runs (third from each of two sets of sweeps) on LTO battery with \SI{3}{A} and \SI{35}{mA} square wave, respectively.}
\label{fmpPlot_Idc_Comparison_run3.pdf}
\end{figure}

In view of the homogeneity between sweeps for each type of measurement (i.e. bz3p66 and bzdcp66 with or without a working square wave), individual representative plots may be used with confidence for visual comparison of the different methods. Figure~\ref{fmpPlot_Idc_Comparison_run3.pdf} contrasts third (final) runs from one set each of bzdcp66 runs with \SI{3}{A} and \SI{35}{mA} working currents.

Note the reduction in impedance magnitude when a significant working signal is applied, particularly in the region below \SI{0.5}{mHz} (e.g. \SI{0.0052}{ohm} vs \SI{0.0118}{ohm} at \SI{0.1}{mHz}), and the elimination of erratic behavior in the low frequency region of the phase plot. Additional shaping of the phase curve is also evident in the region around \SI{10}{mHz} in the presence of a \SI{3}{A} working current. This is not visible when the square wave is effectively `turned off', and implies the availability of additional information that might be important in the development of a realistic ECM for the battery.

\begin{figure}[htbp]
\includegraphics[scale=0.62, trim =100 250 80 250, clip]{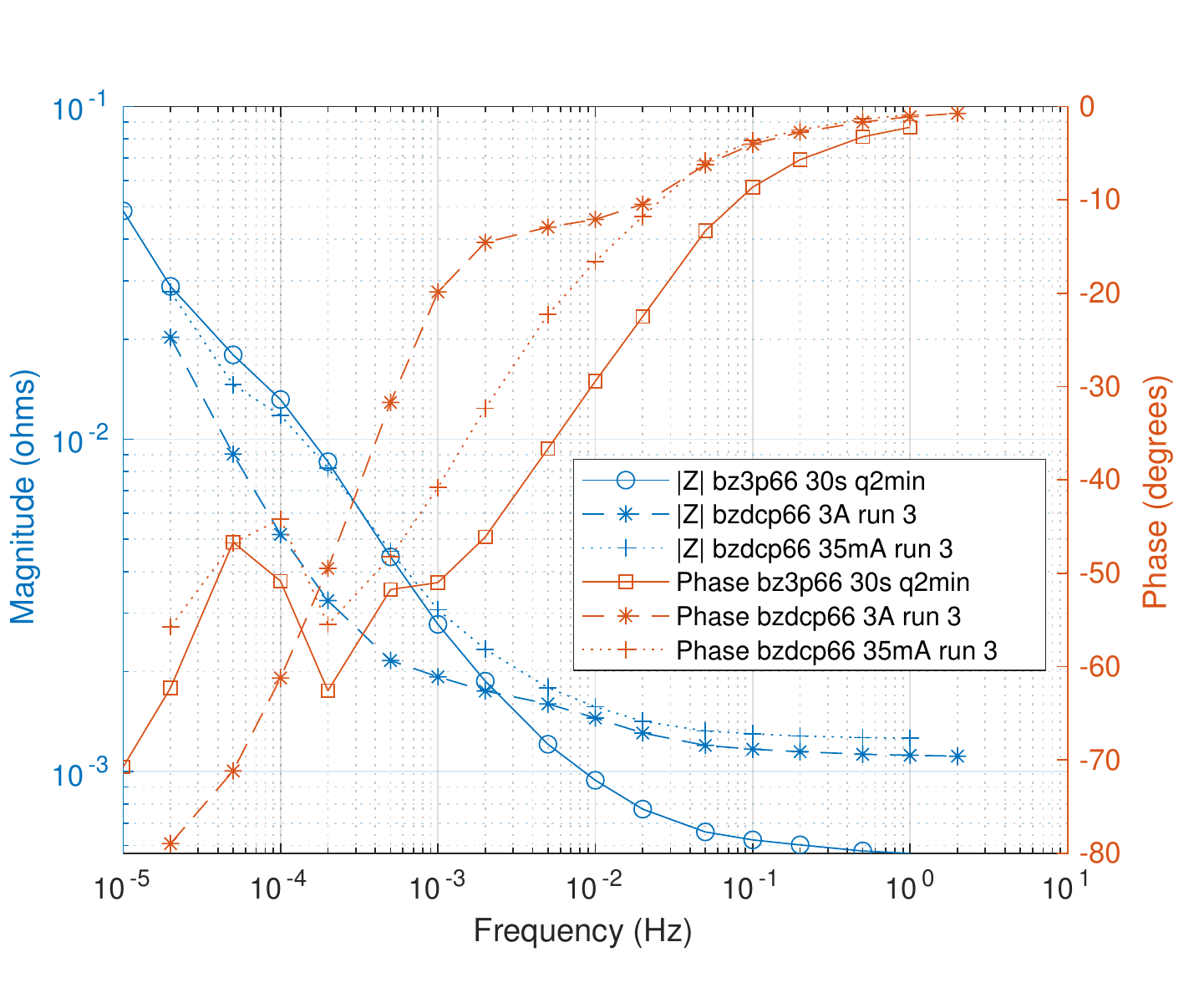}
\caption{Impedance magnitude and phase plots: comparison of bzdcp66 with \SI{3}{A} and \SI{35}{mA}square wave and bz3p66 runs on LTO battery.}
\label{LTO_bz3p_bzdc_Comparison.pdf}
\end{figure}

Finally, Figure~\ref{LTO_bz3p_bzdc_Comparison.pdf} repeats the above with the addition of magnitude and phase information from a representative bz3p66 run (30-second pulse every 2 minutes). The magnitude trace follows that of bzdcp66 with negligible square wave in the low frequency region, but with markedly erratic phase behavior. Most notable, however, is the suppression of impedance magnitude in the higher frequency $ R_s $ region.

\section{Conclusion}

We have presented two novel methods of measuring the impedance of batteries in the presence of
substantial currents. One method time-division multiplexes EIS measurement stimuli with large `working' current bursts, the second method multiplexes in the frequency domain, relying on Fourier transformation to recover the EIS data from `underneath' the large working current.
Both methods break the fundamental limiting tradeoff noted by Budde-Meiwes et al.~\cite{budde-meiwes_influence_2011} between the magnitude of the `working' current that can be drawn during measurement and the lowest attainable frequency in the EIS. We show here measurements to \SI{10}{\micro Hz}, and note that exploratory measurements to below \SI{500}{nanohertz} have been made.

Frequency division multiplexing proves superior in stabilizing the impedance spectrum against impedance drift phenomena. 
That stable impedance agrees most closely with expectations from models~\cite{scott_new_2019, hasan_extending_2020, poihipi_distinguishability_2021, freeborn_fractional-order_2015}.

Substantial charge displacement, better achieved with the frequency domain method, appears to better stabilize the battery impedance, compared with the case of large peak currents achieved with the time domain method.
This implies that it is charge displacement rather than working current value that should be maximized in order to expose the true working impedance of a battery.
This is consistent with the expectation that surface layer processes on the battery electrodes are responsible for the drift. The exact chemical nature of these requires further clarification in the future.

\section*{Acknowledgments}

The authors wish to acknowledge the assistance of WaikatoLink and a Waikato University 
Doctoral Scholarship.

\onecolumn
{\appendix[bz3p66 and bzdcp66 Instructions]
The following directions are shown to users when bz3p66 and bzdcp66 are called with no parameters.

\small
\begin{verbatim}
bz3p66 V6.05 jbs et al, Dec 2020 -> 6 Aug 2021
Battery Z measurement with triphasic pulses via Prologix/Fenrir GPIB-USB & 66332A.
Usage: bz3p66 USB Vmin Vmax Imax dQmax ncyc fmin fmax Xcyc Pf Pw Ip tr baseName [Addr [dftp [ff]]]
where-  USB is the rPi USB address (/dev/ttyUSB0, /dev/ttyACM0, etc);
        Vmin/Vmax are voltage limits (aborts outside this range);
        Imax is the maximum permitted current (-value => only sinks I);
        dQmax is the total charge in Ah that can be sourced or sunk (-val => equal I tones);
        ncyc is the # cycles at fmin (typically 2.01-6.00);
        fmin/fmax are the lowest and highest freqs;
        Xcyc is the # cycles at fmin of data to discard before logging.
        Pf is the frequency of pulse occurences in multitone time, =1/Ttp seconds;
        Pw is the period of the triphasic pulse, in seconds;
        Ip is the peak current of the triphasic pulses;
        tr is the rest period after the triphasic pulse before resuming multitone;
        baseName is the file string to be used;
        Addr is the optional GPIB bus address, def=5.
        dftp is the [path]name of Scott/Farrow dft program (dftp,dvtv,etc);
        ff is the [path]name of the Scott/Finer multitone optimiser program.
Makes a multitone tvi/Z measurement by sourcing current, measuring V & I.
If the USB parameter is set to "skip" the tvi measurement is skipped.
Creates baseName.tvi, basename.log, [.bat, .fmp, [.ffz]] files.
Z optionally computed by calls to dftp [& ff] at each frequency.
.bat file is dft script, fmp has dft's z values, ffz is refined fmp.
Frequencies are a 1-2-5 sequence between fmin and fmax;
if fmax<0 frequencies are read from baseName.frq file, up to 32 freqs.
Requires no drivers, communicates using ++cmd protocol.
Writes complete data, including pulses, to basename.ptvi file.
\end{verbatim}
\hfill
\begin{verbatim}
bzdcp66 ------------  V6.24 jbs Dec 2020 -> Nov 2021
Battery Z measurement with dc, via Prologix/Fenrir GPIB-USB & 66332A, optional DFT.
Usage: bzdcp66 USB Vmin Vmax Imax dQmax ncyc fmin fmax Xcyc Idc fdc baseName [Addr [dftp]]
where-  USB is the rPi USB address (/dev/ttyUSB0, /dev/ttyACM0, etc);
        Vmin/Vmax are voltage limits (aborts outside this range);
        Imax is the maximum multitone current in the cell (if <0 only sinks I);
        dQmax is the total charge in Ah that can be sourced or sunk (-val => equal I tones);
        ncyc is the # cycles at fmin (typically 2.01-6.00);
        fmin/fmax are the lowest and highest freqs;
        Xcyc is the # cycles at fmin of data to discard before logging;
        Idc is the magnitude of the added dc current component, and
        fdc is the frequency of the squarewave at Idc;
        baseName is the file string to be used;
        Addr is the optional GPIB bus address, def=5.
        dftp is the [path]name of Scott/Farrow dft program (dftp,dvtv,etc);
Makes a multitone tvi/Z measurement by sourcing current, measuring V & I.
If the USB parameter is set to "skip" the tvi measurement is skipped.
NB: If Imax<0 battery is discharging, so dQmax boundary checking is ignored.
Creates baseName.tvi, basename.log, [.bat, .fmp] files.
Z optionally computed by calls to dftp at each frequency.
The .bat file is dft script, fmp has dft's z values.
Frequencies are a 1-2-5 sequence between fmin and fmax;
if fmax<0, frequencies are read from baseName.frq file, up to 32 freqs.
fdc should be chosen so that none of its harmonics clash with multitones.
Requires no drivers, communicates using ++cmd protocol.
Measures for (ncyc+Xcyc)/fmin seconds, then does dft calls.
Corrects for 1/2 LSB DAC error in 66332.
\end{verbatim}
}

\begin{multicols}{2}

\bibliographystyle{IEEEtran}
\bibliography{IEEEabrv,references}

%
%
%
%

\vfill
\end{multicols}
\end{document}